\documentclass[aps,pre,twocolumn,floats]{revtex4}
\usepackage{epsfig}
\begin{document}

\newcommand{\tbox}[1]{\mbox{\tiny #1}}
\newcommand{\half}{\mbox{\small $\frac{1}{2}$}}
\newcommand{\mbf}[1]{{\mathbf #1}}


\title{Short time decay of the Loschmidt echo}
\author{Diego A. Wisniacki}
\date{\today}
\email[]{wisniacki@df.uba.ar}
\affiliation{Departamento de F\'{\i}sica, "J. J. Giambiagi," 
FCEN, UBA, Pabell\'on 1, Ciudad Universitaria, 1428 Buenos Aires, Argentina.}


\begin{abstract}
The Loschmidt echo measures the sensitivity to perturbations
of quantum evolutions. We study its short time decay in 
classically chaotic systems. 
Using perturbation theory and throwing out all correlation 
imposed by the initial state and the perturbation, we show 
that the characteristic time of this regime is well
described by the inverse of the width of the local density 
of states. This result is illustrated and discussed in a 
numerical study in a 2-dimensional chaotic billiard system 
perturbed by various contour deformations and
using different types of initial conditions. 
Moreover, the influence to the short time decay of 
sub-Planck structures developed by time evolution 
is also investigated.
\end{abstract}

\maketitle

\section{INTRODUCTION}

Quantum irreversibility studies have become a very active research 
topic due to a direct connection with quantum computers and 
mesoscopics physics \cite{qcomputer,mesos}.
The natural quantity for these investigations has been introduced
by Peres in his seminal paper of 1984 \cite{peres}. 
Called later as Loschmidt echo (LE) or fidelity, it 
measures the ability of a system to return to an initial state 
$|\phi \rangle$
after a forward evolution with a Hamiltonian $H_{0}$ followed by 
an imperfect reversal evolution with a perturbed Hamiltonian 
$H=H_{0}+ \delta x H'$ ($\delta x$ parameterize the strength of the
perturbation). Thus, it is given by
\begin{equation}
M(t)=|\left\langle \phi \right| \exp [{\rm i} H t ]
\exp [-{\rm i} H_{0} t ]\left| \phi \right\rangle |^{2}
\label{Eq-overlap}
\end{equation}
(throughout the paper  $\hbar$ is set equal to 1). The LE compares 
the evolution of an initial state
with slightly different Hamiltonians and can distinguish regular and
chaotic classical dynamics \cite{peres,jalabert,jaquod1}.
  
The LE was recently studied in various chaotic systems using 
several approaches \cite{jalabert, jaquod2, tomsovic, 
fernando1, prosen, wisniacki1,fernando2,wisniacki2}.
However, few types of decay were discussed in the literature. 
For a very short time, it is straightforward to show that the LE 
has a parabolic behavior $M(t)=1-\delta x^2 (\Delta H')^2 t^2$, 
with $(\Delta H')^2=[\langle \phi|H'^2 |\phi \rangle-
\langle \phi|H'|\phi \rangle^2]$.
This decay is better resembled by the Gaussian function 
$\exp[-(t/\tau)^2]$, with characteristic time $\tau=1/(\Delta H' \delta x)$. 
Though this regime has experimental
relevance \cite{LEexp}, it has not been extensively taken
into account. 

After this short time decay, a crossover to a perturbation 
dependent regime was predicted and numerically observed 
\cite{jalabert,tomsovic,jaquod2,fernando1, wisniacki1}. 
For very small $\delta x$, in which a typical matrix element 
$U$ of the perturbation is smaller than the mean level 
spacing $\Delta$, the decay 
is always Gaussian until $M(t)$ reaches its asymptotic values 
$M(t \rightarrow \infty) \equiv M_{\infty}$. 
If $U>\Delta$, this regime has an exponential decay 
$\exp(- \Gamma t)$, with $\Gamma$ the width of the local
density of states (LDOS). This is usually called Fermi golden 
rule regime (FGR). When $\Gamma > \lambda$, with $\lambda$ the
mean Lyapunov exponent 
of the classical system, a perturbation independent regime is
observed. In this case, the decay rate is given by $\lambda$. 
Finally, if $\Gamma$ exceeds the bandwidth of the perturbation,
the LE has a Gaussian decay.
 
The properties of the initial state play an important
role in the behavior of the LE \cite{jaquod2,wisniacki2}. 
This point can be relevant to observe the 
mentioned regimes. 
For example, the Lyapunov regime is not displayed if 
the initial state is an eigenfunction of the unperturbed/perturbed 
Hamiltonian \cite{wisniacki2}. Localized
wave packets are needed to observe this regime. 
On the other hand, Zurek has recently
stated that dynamical evolution causes that these states 
develop a sub-Planck structures in phase space, 
and he predicts that these structures enhance their 
sensitivity to perturbations \cite{zurek}.

In this article we are mainly interested in the 
short time decay of the LE. Disregarding system 
specific features and the correlations imposed by
the characteristics of initial state, we show
via perturbation theory that $\tau^{-1}$ is given
by the width $\Gamma$ of the LDOS.
In order to see the validity of this result in a realistic model, 
we study the
characteristic time $\tau$ in a paradigmatic
model of quantum chaos, a 2-D chaotic billiard perturbed 
by a contour deformation. 
We regard the influences of the 
characteristics of the perturbation and the initial condition to 
this important regime. Therefore, we consider different perturbations 
and various types of initial conditions in our numerical study.
We find that some perturbations, which we call {\it generic}, 
destroy  correlations imposed 
by the initial condition giving that $\tau^{-1} = \Gamma$. 
Nevetheless, other perturbations do not act in that way 
and this fact produces a slower decay with $\tau^{-1} < \Gamma$.
In this context, we discuss the influence
of an initial time evolution of the wave packet and the corresponding
developed structures in phase space in the short time decay of the LE.
In fact, it is shown that an initial dynamical evolution helps to erase
the mentioned correlations, with the effect of increasing $\tau^{-1}$.
If the initial evolved time is smaller than the Ehrenfest time, the enhanced decay 
is described entirely by the classical streaching arround the unstable manifold
given by the Lyapunov exponent \cite{jaquodsubplanck}.
But for greater evolved times, the quantum interference lead a developed 
sub-planck structure in phase space and this yield that the decay continuos 
growing. That is, the sensitivity to perturbations is also enhanced in this case (as stated in
Ref. \cite{zurek}). However , if the perturbation produces a decay with $\tau^{-1}=\Gamma$
the developed structures in phase space do not influence the short time decay.  

The paper has the following structure. Sec. \ref{model} is devoted to describe the 
model system and the various shape deformations that we have considered. 
The paper is self-contained with the inclusion of the
shape parameter Hamiltonian expansion for a 2-D quantum 
billiards developed in Ref. \cite{wisniacki3}. 
In Sec. \ref{shorttime}, the characteristic time
$\tau$ is related
to the width of the LDOS using perturbation theory.  
Then, in Sec. \ref{echo}
we present the numerical results. We study the LE for several 
types of initial condition.  Starting with the 
simplest case when the initial state is an eigenfunction of $H_0$, we 
follow with Gaussian wave packets. Finally the initial conditions are 
the evolved Gaussian wave packets in order to study the prediction of
Ref. \cite{zurek}. In Sec. \ref{finalremarks} we make some final remarks. 

\section{MODEL SYSTEM: DEFORMED STADIUM BILLIARDS}

\begin{figure}[h]
\centerline{\epsfig{figure=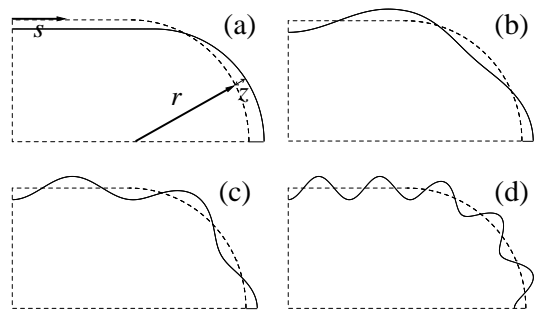,width=7cm}}
\vspace{.1in}
\caption{Schematic figure of the system and the various shape deformations. 
Desymmetrised stadium billiard is plotted with dashed lines. On continuous lines,
deformations of the billiard are shown. The curvilinear coordinates 
used to describe the deformations [Eq. 2] is also included.}
\label{system}
\end{figure}

We use the desymmetrized Bunimovich stadium billiard as a model system to 
explore the behavior of the LE \cite{wisniacki1}. 
This paradigmatic system
is fully chaotic and has great theoretical
and experimental relevance 
\cite{cit-bunimovich,cit-marcus-dephas-dot,cit-marcus-def-dot}.
It consists of a free particle inside a 2-dimensional planar region whose boundary  
${\cal C}$ is shown in Fig. ~\ref{system} with dashed lines. The radius
$r$ is taken equal to unity and the enclosed area is $1+\pi /4$. 
 
The system is perturbed by boundary deformations which preserves the area of 
the billiard. Deformations with different characteristics
are chosen in order to understand their influences 
in the short time decay of the LE. Fig.  ~\ref{system} shows the shape deformations
that have been considered. The changes of the boundary are parameterized by 
\begin{equation}
{\bf r} (s,\delta x )= {\bf r_{0}} (s)+z(s,\delta x )\;{\bf n}\;,
\label{borde}
\end{equation} 
with $s$ along ${\cal C}$, ${\bf r_{0}} (s)$ the parametric equation 
for ${\cal C}$, and 
${\bf n}$ the outward normal unit vector to ${\cal C}$ at ${\bf r}_{0}(s)$
(see Fig. ~\ref{system}(a)).
Case (a), shown in Fig.  ~\ref{system}, is well described in
Ref. \cite{wisniacki3}.
For deformation (b)-(d),  
$z(s,\delta x)=\alpha \delta x \cos(\pi N s /P)$ with
$\alpha=0.42$, $P=1+\pi/2$ and N=3, 5 and 10. 
$\alpha$ is chosen in order that the
width of the LDOS of all deformations are aproximately
equal in the studied range of $\delta x$ . 
As we shall see in the next section, the width of the LDOS is a measure 
of the magnitude of a perturbation.  

To solve Eq. (1), eigenvalues and
eigenfunctions of perturbed and unperturbed system
are needed.
They are obtained using a Hamiltonian expansion of 
deformed billiards which has been recently developed
\cite{wisniacki3}.  For $\delta x \leq 1/k_{0}$
($k_{0}$ the mean wave number of the region under study), the
eigenenergies  and eigenfunctions of the deformed billiard, are 
connected to the ones of the stadium by the linearized Hamiltonian
expressed in the basis of eigenstates at $\delta x =0$ (from now on
we will call $\phi_{\mu }$ to these eigenstates and 
$E_{\mu }$ the respective eigenenergies) ,
\begin{equation} 
H_{\mu\nu}(\delta x )\simeq H_{\mu\nu}(0)+ \delta x \; H'_{\mu\nu},
\label{exp}
\end{equation}
with $H_{\mu\nu}(0)=E_{\mu} \delta_{\mu \nu}$ and
\[
H'_{\mu\nu}=-{\rm Cf}_{\mu\nu}\;A_{\mu\nu}
\oint_{{\zeta}}z'(s)\; 
\frac{\partial \phi_{\mu}}{\partial {\bf n}}
\frac{\partial \phi_{\nu}}{\partial {\bf n}}ds.
\]
The eigenfunctions and eigenenergies
at $\delta x=0$ are obtained using the 
scaling method \cite{cit-Vergini-Saraceno}.
The integral above could be viewed as an inner product among the wave
functions $\frac{\partial \phi _{\mu }}{\partial {\bf n}}$ defined over $%
{\cal C}$. This relation defines an effective Hilbert space in a window $%
\Delta k \approx$ Perimeter/Area \cite{wisniacki3}. The cut-off function
${\rm Cf }_{\mu \nu }=\exp \left[ -2\;(k_{\mu }^{2}-k_{\nu
}^{2})^{2}/(k_{0} \Delta k)^{2}\right] $ restricts the effect of the
perturbation to states in this energy shell of width  $k_{0} \Delta k$. 
It allows to deal with a basis of finite dimension with wave numbers
around the mean value $k_{0}$, restricting to a particular region 
$\Delta k$  of interest. We are considering $k_{0}=100$ and $2m=1$
($m$ the mass of the particle)
in all the numerical
calculations presented above. 

\label{model}
\section{SHORT TIME DECAY AND THE LOCAL DENSITY OF STATES}

Our aim is to characterize the short time decay of the LE.
As mentioned in the introduction, a simple calculation led  the short
time decay to depend on the initial state and on the perturbation \cite{peres}. 
We want to relate the characteristic time $\tau$ of the 
short time decay with some general properties of the perturbation.
 
The influence of a perturbation over a quantum system could be described by
the local density of states (LDOS). 
The LDOS of an unperturbed eigenstate $\phi_{\mu}$ is defined as 
\begin{equation}
\rho_{\mu}(E,\delta x)=\sum_{\nu}{|\langle \phi_{\nu}(\delta x) | {\phi_{\mu}}  \rangle|^2 
\delta[E- (E_{\nu}(\delta x)-E_{\mu})]},
\end{equation}
with $E_{\nu}(\delta x)$ and $\phi_{\nu}(\delta x)$ the energy and eigenfunction 
of the perturbed Hamiltonian [Eq. (\ref{exp})]. 
This function shows how the unperturbed 
states are coupled to the perturbed ones. 
Because we are not interested in a particular state, an implicit average
over the unperturbed state $\mu$ is considered from now on. We have chosen the width (dispersion)
\begin{equation}
\Gamma(\delta x)=\sqrt{\sum_{\nu}{\rho_{\mu}(E_{\nu}(\delta x)-E_{\mu},\delta x) (E_{\nu}(\delta x)-
E_{\mu})^2}},
\label{defdis}
\end{equation}
as a practical measure of this distribution .

The LDOS exhibits various regimes as a function of the strength $\delta x$ 
\cite{dcohenannal,dcohenpara}. As we shall see, the perturbative regime 
is relevant for our study. Perturbation theory (PT) gives the following first order
expression for the LDOS
\begin{equation}
\rho_{\rm PT}(E,\delta x)=\delta(E) + 
\frac{|H'_{\mu \nu}|^2 \delta x^2}{[E_{\nu}(\delta x)-E_{\mu})]^2}
\delta[E-(E_{\nu}(\delta x)-E_{\mu})].
\label{rhopt}
\end{equation} 

Using the definition of the width [Eq. (\ref{defdis})] and Eq. (\ref{rhopt}) it is 
straightforward to show  
\[
\Gamma_{\rm PT}(\delta x)=\delta x \sqrt{\sum_{\nu}|H'_{\mu \nu}|^2}. 
\]
This expression works very well for all the perturbations of Fig \ref{system}
with strength $\delta x \le 1/k_0$.

With these ingredients in mind, let us consider the short time decay of the LE.
As pointed out previously, it is given by $M(t)=\exp[-(t/\tau)^2]$ 
with $\tau^{-2}=(\langle \phi| H'^2 |\phi \rangle-
\langle \phi|H'|\phi \rangle^2) \delta x^2$. Let be  
$|\phi \rangle = \sum{ a_{\mu} \phi_{\mu } }$ the initial state , so 

\begin{equation}
\tau^{-2}=[\sum{a_{\mu} H'_{\mu i } H'_{i \nu} a_{\nu}}-
(\sum{a_{\mu} H'_{\mu \nu } a_{\nu}})^2] \delta x^2
\label{dispersion}
\end{equation}

\begin{figure}[h]
\centerline{\epsfig{figure=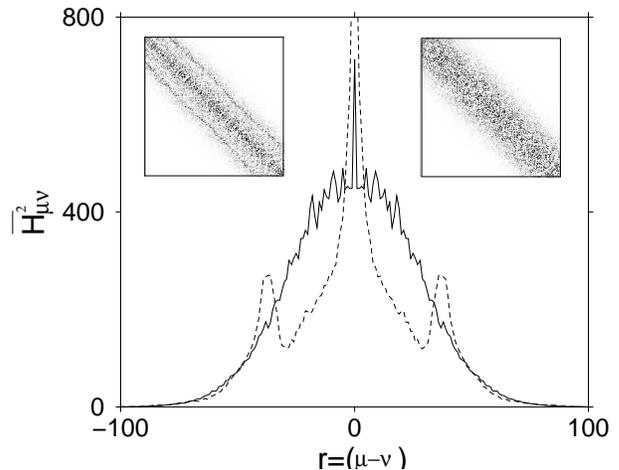,width=8cm}}
\vspace{.1in}
\caption{Mean value of the matrix elements $H'^2_{\mu \nu }$ as a function of 
$r=\mu-\nu$
for the deformations (a) (dotted line) and (d) (full line). 
The peak for the deformation
(a) is due to the bouncing ball orbits. Insets:  Image of a piece of the 
perturbation matrix
$|H'_{\mu\nu}|$ shown as a density plot. The left plot corresponds to the 
perturbation label as (a)  and the right plot corresponds to
perturbation label as (d). Clearly, there is a structure in the perturbation
matrix for pertubation (a) when compared to perturbation (d).}
\label{Hprima}
\end{figure}

The perturbation matrix $H'_{\nu \mu}$ is a banded matrix due to the cut-off function
${\rm Cf }_{\mu \nu }$ (see the insets of Fig. \ref{Hprima}). This band structure
is quite generic in realistic systems due to the finite range interaction of
unperturbed states \cite{band1,band2}. 
Inside the band, the matrix elements $H'_{\mu\nu}$ are highly fluctuating numbers. 
At first sight if we ignore the system specific features, we can do the 
diagonal approximation of Eq. \ref{dispersion}, resulting  
\begin{equation}
\tau^{-2}=[\sum{|a_{\mu}|^2 H'^2_{\mu \nu }}] \delta x^2.
\label{tauaprox}
\end{equation}
This approximation is also valid for the case in which the complex numbers
$a_{\mu}$ behave randomly. Finally, if we consider that the $|a_{\mu}|^2$ are nearly
constant inside the band,
 
\begin{equation}
\tau^{-2} \approx [\sum{ H'^2_{\mu \nu }}] \delta x^2 = \Gamma^2_{\rm PT}(\delta x).
\label{ptresult}
\end{equation} 

This result is also valid if $a_{\mu}=\delta_{\mu \nu}$, assuming that we are averaging
over several initial conditions. Note that we have obtained Eq. (\ref{ptresult})
throwing out all the correlations imposed by the perturbation and the wave amplitudes. 
In the next section we will see in what manner these correlations influence the decay.   

Each perturbation is characterized by the structure and correlations of the matrix elements
$H'_{\mu\nu}$.
As an example,  Fig. \ref{Hprima} shows the behavior of the mean value of 
$H'^2_{\mu\nu}$ as a function of the distance $r = \nu-\mu$ for perturbations (a) and (d).
This function is usually called band profile.
The cases (b) and (c) are not plotted because their band profile are qualitative equal 
to case (d). 
Note that perturbation (a) is clearly {\it non-generic} due to the two important peaks 
at $|r| \sim 25$. This non-universality is introduced by  the fact that perturbation (a)
does not connect the bouncing ball states with generic states.
In the insets of Fig. ~\ref{Hprima} a density plot of these matrix are shown for 
deformations (a) and (d).  
We will see in the next section that there are correlations
between the matrix elements of the 
perturbation which are not exposed in the band profile but
have an important influence to the short time decay.

\label{shorttime}
\section{NUMERICAL RESULTS}

In this section we study numerically the behavior of the short time decay 
of the LE in the
Bunimovich stadium billiard perturbed by the contour deformation presented
in Sec. \ref{system}. We consider  different types of initial conditions: 
Eigenfuction of $H_{0}$,
Gaussian wave packets and evolved Gaussian wave packets . We would like to see
the range of validity of Eq. 
\ref{ptresult} for our particular system.

\label{echo}

\subsection{Eigenfuctions of $H_{0}$}

The simplest case of the LE is when the initial state
of Eq. \ref{Eq-overlap} is
an eigenstate of ${\cal H}_{0}$. In this case, the LE
is directly related with the Fourier transform (FT) of
the LDOS. Then, $M(t)$ is the so called 
survival probability \cite{wisniacki2,Cohen-Heller}, defined as
\begin{equation} 
P(t) =|\left\langle \phi_{\mu} \right| \exp [{\rm i} H t ] 
\left| \phi_{\mu} \right\rangle |^{2}=|FT[\rho_{\mu}(E,\delta x)]|^2   
\end{equation}

\begin{figure}[h]
\centerline{\epsfig{figure=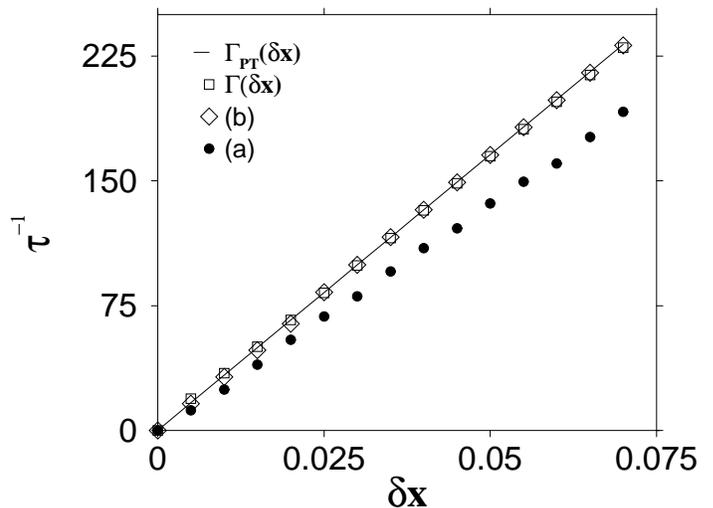,width=7cm,angle=-90}}
\vspace{.1in}
\caption{$\tau^{-1}$ of the survival probability as a function of $\delta x$ when the system is perturbed 
by deformations (a) and (b). The width of the LDOS $\Gamma(\delta x)$ and its perturbative evaluation
$\Gamma_{PT} (\delta x)$  calculated using Eq. (6) are also plotted.}
\label{tausp}
\end{figure}

As we saw in the previous section, we expect that $\tau$ for $\bar{P}(t)$ 
is well described by Eq. \ref{ptresult}. $\bar{P}(t)$ is the mean value 
of the survival probability over several initial states. 
We have numerically observed that this is 
the case for the perturbations (b-d). These results are
shown in Fig. \ref{tausp}. Note that $\tau^{-1}$ for perturbations (c) and (d)
are not plotted because the results are the same as (b). 
The widths $\Gamma (\delta x)$ and 
$\Gamma_{PT}(\delta x)$ are equal
for all the perturbations.  For perturbation
(a), $\tau^{-1} \approx 0.85 \Gamma(\delta x)$ which is directly related to the
structure of the perturbation matrix (see Fig. \ref{Hprima}) imposed by 
the bouncing ball states \cite{wisniacki1}. Similar attenuation was observed
in Ref. \cite{wisniacki1}
for perturbation (a) in the FGR regime. In this case, the decay rate is
given by $0.5 \Gamma(\delta x)$ insted of the expected value $\Gamma(\delta x)$. 

\begin{figure}[h]
\centerline{\epsfig{figure=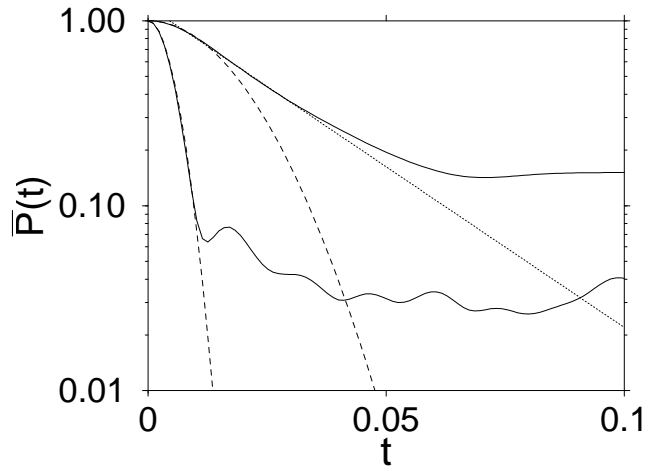,width=8.5cm,angle=0}}
\vspace{.1in}
\caption{$\bar{P}(t)$ for the desymmetrized stadium billiard perturbed by deformation (d) of Fig. 1.
The values of the strength of the perturbation is $\delta x=0.015$ for the top curve and 
0.05 for the bottom curve.
The dashed lines 
correspond to the Gaussian decay $\exp[-(\Gamma(\delta x) t)^2]$ and 
the exponential decay $\exp[-(\Gamma(\delta x) t)]$ is plotted
with a dotted line.}
\label{P(t)}
\end{figure}

Fig. \ref{P(t)}
summarizes the behavior  of $\bar{P}(t)$ in the desymmetrized stadium billiard. 
We have taken an average over 100 initial states.
In this figure, 
the results for perturbation (d) of Fig. 1 are shown. Other perturbations
led to the same qualitative results.
For a small perturbation strength (top curve of Fig. \ref{P(t)}) we observe the Gaussian short 
time decay and after
that an exponential decay with a decay rate given by the width of the LDOS 
\cite{jaquod1,wisniacki1}. For large 
perturbation strength (bottom curve of Fig. \ref{P(t)}) the 
exponential decay with decay rate given by $\Gamma(\delta x)$ is not  observed . 
This is due to the constraint 
imposed by the asymptotic value $M_{\infty}$. In this case  the decay is completely 
Gaussian. An important point is that the asymtotic 
$M_{\infty} \approx \Delta/\Gamma(\delta x)$.

\subsection{Localized Gaussian wave packets}

The decay of the LE for localized Gaussian wave packets has been
widely studied in the literature. Most of the previous
works consider this case
\cite{jalabert,jaquod2,wisniacki1,fernando1,fernando2}.
The predicted crossover from a perturbation dependent regime to
the Lyapunov regime has been shown  
for these classically adapted initial conditions.

\begin{figure}[h]
\centerline{\epsfig{figure=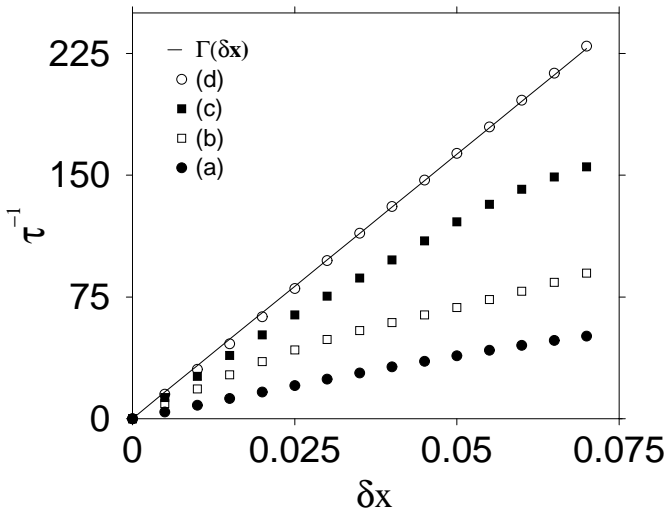,width=7cm,angle=-90}}
\vspace{.1in}
\caption{$\tau^{-1}$ of $\bar{M}(t)$ as a function of $\delta x$ when the system is perturbed
by deformations (a), (b), (c) 
and (d). The initial states are the Gaussian wave packets of Eq. (11). 
$\Gamma (\delta x)$ is also  plotted in solid line.}
\label{taugau}
\end{figure}

We discuss here the short time decay of  the LE for this particular
type initial conditions in the stadium billiard perturbed by  
deformations presented in Fig. 1. 
We compute $M(t)$ for initial Gaussian wave packets,
\begin{equation}
\phi(\vec{r})=(\pi \sigma^2)^{1/2} \exp{[i \vec{p}_0.(\vec{r}-\vec{r}_0)-
i|\vec{r}-\vec{r}_0|^2 /\sigma^2]}
\label{gau}
\end{equation}
with $|p_{0}|=k_{0}=100$ and $\sigma=0.16$. An average over 50 initial states was taken.
The direction of the momentum  $\vec{p}_0/|\vec{p}_0|$ and
the center of the wave packet $\vec{r}_0$ are chosen randomly. 
As expected, the short-time decay is well described by a Gaussian
function $\exp[-(t/\tau)^2]$. Fig. \ref{taugau} shows the behavior of
$\tau^{-1}$ as a function of the strength $\delta x$ for
all the perturbations under study. 
This type of initial conditions imposes correlations so that 
the second sum of right hand side of
Eq. (\ref{dispersion}) does not vanish.
The non-universalities of each perturbation are
clearly exposed in $\tau$. Note that these differences are not seen
in the width of the LDOS nor in band profile of the matrix $H'_{\mu\nu}$.    
We have considered perturbations with greater
number of oscillations of the boundary ($N >10$) and we find 
$\tau=1/\Gamma(\delta x)$ for all of them \cite{wisniacki4}.  

\subsection{Evolved Gaussian wave packets}

In a recent paper \cite{zurek}, Zurek shows that dynamical evolution
of initial Gaussian wave packets in classically chaotic systems,
produces finer and finer phase space structures which saturates
after the Ehrenfest time with a sub-Planck scale.
More important, he predict that this sub-Planck structures
enhances the sensitivity of a quantum state
to an external perturbation. A numerical study in
a time dependent one dimensional model agreed with this assertion
\cite{karkuszewski}, but other 
studies reached opposite conclusions \cite{jaquodsubplanck,jordan}.
In Ref. \cite{jaquodsubplanck} it is showed an enhanced decay
of the LE of evolved wave packets but this 
acceleration is fully described by the classical Lyapunov exponent
and it is not due to the sub-Planck structures. More specifically,
it is pointed out that
the characteristic time $\tau (T_{0})$ of the short 
time decay for a wave packet that has been evolved a time $T_{0}$ is
given by
\begin{equation}
\tau^{-1}(T_{0})=\tau^{-1}(0) \exp(\frac{\lambda (T_{0}-T_{c})}{2}),
\label{shorttau}
\end{equation}    
with  $\tau(0)$ the characteristic time
of the short time decay for initial states that have not been evolved,
$\lambda$ the Lyapunov exponent and $T_{c}$ the mean time for the first
 collision with the boundary.

\begin{figure}[h]
\vspace{.1in}
\centerline{\epsfig{figure=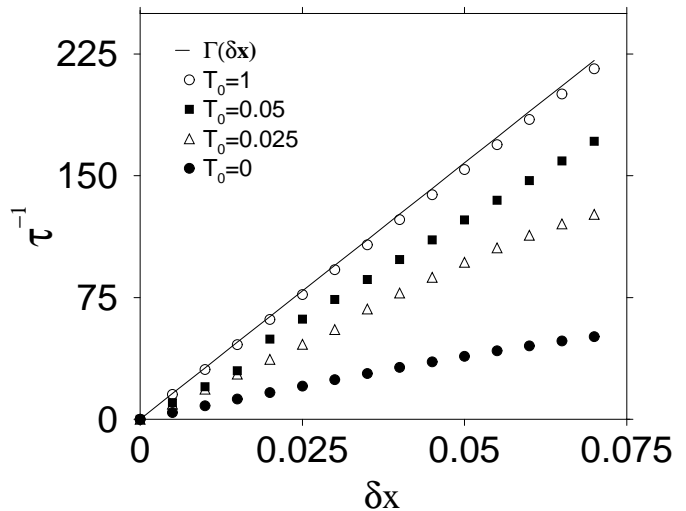,width=7cm,angle=-90}}
\caption{$\tau^{-1}$ of $\bar{M}(t)$ as a function of $\delta x$ for perturbations (a).
The initial states are Gaussian wave packets that have been evolved a time $T_{0}$. 
$\Gamma(\delta x)$ is also plotted.}
\label{tauzurek}
\end{figure}

We consider the influence of the developed structures in phase space
in the short time decay of the LE. So, $M(t)$ is computed for the 
same initial conditions [Eq. (\ref{gau})] of the previous section but 
an unperturbed initial evolution during a time $T_{0}$ is applied.
Fig. \ref{tauzurek} shows  $\tau^{-1}(T_{0})$ as a function of the strength $\delta x$ 
for perturbation (a) and with $T_{0}=0,0.025, 0.05,$ and 1. 
Note that $\tau^{-1}(T_{0})$ increases with larger $T_{0}$, for all perturbation strenghts. 
This fact clearly points out that
an initial evolution enhances the sensitivity to this particular perturbation.
We have observed that for $T_{0} > 0.4$, $\tau^{-1}$ converges to the width 
of the LDOS. Same behavior is shown when the system is perturbed by deformations (b) and (c). 
However, when the perturbation destroys the correlations imposed by the initial wave packet 
which implies that $\tau=1/\Gamma$, the short time decay is not affected by an initial 
evolution.  
 
\begin{figure}[h]
\vspace{.1in}
\centerline{\epsfig{figure=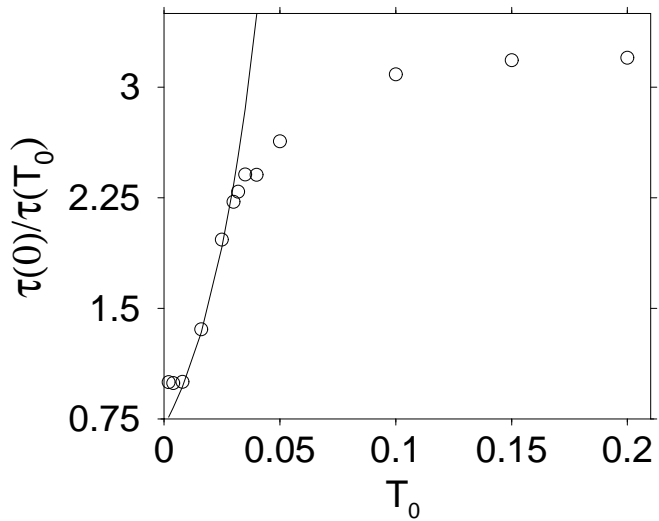,width=7cm,angle=-90}}
\caption{$\tau^{-1}(T_{0})/\tau^{-1}(0)$ as a function of the preparation 
time $T_{0}$ 
for strenght $\delta x=0.04$. The stadium is perturbed by deformation (a).
The prediction of Ref. \cite{jaquodsubplanck} (Eq. \ref{shorttau}) is ploted
in solid line.}
\label{tauzurek2}
\end{figure}

In order to see if these increments are fully described
by Eq. \ref{shorttau} and due to that has classical nature, 
in Fig.\ref{tauzurek2} 
it is showed the behavior of
$\tau^{-1}(T_{0})/\tau^{-1}(0)$ for several preparation time $T_{0}$ 
for perturbation (a) with strenght $\delta x=0.04$. 
It is clearly observed that 
Eq. \ref{shorttau} works well for $T_{0}<0.025$. Note that the Ehrenfest time
$T_{E}=0.025$. So, the enhacement of the short time decay is fully explained with 
the classical inestability given by the Lyapunov exponent for evolved times 
$T_{0}$ smaller than the Ehrenfest time. However, for larger times 
$\tau^{-1}(T_{0})$ is also growing. 
 
A qualitative picture of acceleration of the short time decay for evolved
states is the following.
Before the Ehrenfest time the wave packet is streaching
around an unstable manifold and just after that times starts the quantum 
interference which lead a sub-Planck structures. At that times a small
part of the Wigner function presents a sub-Planck structure. This region
grows with time and it seems to be the reason of the accelerating decay.
Note that for the saturation times  the
sub-Planck structure is all arround the available phase space.   
    
\section{FINAL REMARKS} 

We have studied the short time decay of the LE in
a 2-D chaotic billiard. The system was perturbed by
a contour deformation. Different perturbations
were considered in order to develop the influences of 
their characteristics in the behavior of the LE. Moreover, 
several types of initial conditions have been used and 
how they affect the LE have been examined. 

Our findings are the following. For non-localized initial states
and if the system is perturbed by a {\it generic} deformation,
the characteristic time $\tau$ of the short time Gaussian decay is
given by the inverse of the width $\Gamma$ of the LDOS. If semiclassical
features are exposed in the matrix elements of the perturbation, 
we have obtained that 
$\tau^{-1}<\Gamma$. For highly localized initial states, cross correlation 
between wave amplitudes are important and this is exposed with the fact 
that $\tau^{-1}<\Gamma$. When the perturbation destroys such correlations,
the characteristic time $\tau^{-1}$ exhibits its maximum value $\Gamma$.

We have discussed the prediction of Zurek \cite{zurek} which stated 
that an initial dynamical evolution of semiclassical wave packets
lead a sub-Planck structures in phase space and this enhances its 
sensitivity to perturbation.
We found that in the cases in which the perturbation does not destroy
the correlation mentioned before an accelerated decay is observed. 
As a function of the preparation time $T_{0}$, we have observed two regimes.
For $T_{0}$ smaller
than the Ehrenfest time, the enhanced decay 
is described entirely by the classical Lyapunov exponent as pointed out
in Ref. \cite{jaquodsubplanck}. However, for larger $T_{0}$ in which 
the quantum interference lead the sub-Planck structures the 
enhancement of decay is also observed.

A final point is worth commenting. We have shown that the short time decay
of the LE has a Gaussian behavior and for certain perturbations the 
characteristic time is given by the width of the LDOS. The LDOS of some systems
\cite{Cohen-Heller,dcohenpara,wisniacki2,note} presents a region in which its width
is independent of the perturbation. 
We note that these results could be of importance for the understanding of
the measure of the LE in recent  NMR experiments \cite{LEexp}. Although 
that system consists of {\it many} interacting nuclear spins, the results are
in accordance with the former. That is, due to imperfections
in the reversed evolution the LE in the MNR experiment shown a Gaussian
attenuation, 
$M(t)=\exp[(t/\tau)^2]$,
with $\tau$ depending on the small non-inverted interaction (which corresponds to the 
perturbation $\delta x H'$ of Eq. \ref{Eq-overlap}), and in a range of small perturbation,
 the characteristic time $\tau$ do not depend on it.  
\label{finalremarks}

\section*{ACKNOWLEDGMENTS}

This research was partially supported by CONICET and ECOS-SeTCIP. 
I want to thank Fernando Cucchietti, Philippe Jacquod, Horacio Pastawski, 
Juan Pablo Paz, Fabricio Toscano and Eduardo Vergini for very useful 
discussions and comments.



\begin{thebibliography}{99}
\bibitem{qcomputer}
M. A. Nielsen and I. L. Chuang, {\it Quantum Computation and 
Quantum Information} (Cambridge University Press, Cambridge, 2000). 

\bibitem{mesos}
Y. Imry, {\it Introduction to mesoscopic Physics} 
(Oxford, New York, 1997)

\bibitem{peres}
A. Peres, Phys. Rev. A {\bf 30}, 1610 (1984).

\bibitem{jalabert}
R.A. Jalabert and H.M. Pastawski, Phys. Rev. Lett. {\bf 86}, 2490 (2001).

\bibitem{jaquod1}
Ph. Jacquod, I. Adagideli and C.W.J. Beenakker,
nlin.CD/0206160.

\bibitem{jaquod2}
Ph. Jacquod, P.G. Silvestrov and C.W.J. Beenakker,
Phys. Rev. E {\bf 64}, 055203 (2001).

\bibitem{tomsovic}
N.R. Cerruti and S. Tomsovic, Phys. Rev. Lett. {\bf 88}, 054103 (2002).

\bibitem{fernando1}
F.M. Cucchietti,  H. M. Pastawski and D. A. Wisniacki,
Phys. Rev. E  {\bf 65} 045206(R) (2002).

\bibitem{prosen}
T.~Prosen and M.~Znidaric, J. Phys. A {\bf 35} ,1455 (2002).

\bibitem{wisniacki1}
D.A. Wisniacki,  E. G. Vergini, H. M. Pastawski and F. M. Cucchietti, 
Phys. Rev. E  {\bf 65} 055206 (R) (2002).

\bibitem{fernando2}
F.M. Cucchietti, C. H. Lewenkopf, E. R. Mucciolo, H. M. Pastawski and 
R. O. Vallejos, Phys. Rev. E  {\bf 65} 046209 (2002).

\bibitem{wisniacki2}
D.A. Wisniacki and D. Cohen, Phys. Rev. E 66, 046209 (2002)

\bibitem{LEexp}
H. M. Pastawski, P. R. Levstein, G. Usaj, J. Raya and J. Hirschinger,
Physica A, {\bf 283} 166 (2000).

\bibitem{zurek}  W. H. Zurek, Nature {\bf 412}, 712 (2001).

\bibitem{jaquodsubplanck} Ph. Jacquod, I. Adaglideli and C.W.J. Beenakker,
Phys. Rev. Lett. 89, 154103 (2002).

\bibitem{wisniacki3}  D. A. Wisniacki and E. Vergini, Phys. Rev. E {\bf %
59}, 6579 (1999).

\bibitem{cit-bunimovich}  L. A. Bunimovich, Funct. Anal. Appl. {\bf 8} 
254 (1974).

\bibitem{cit-marcus-dephas-dot}  A. G Huibers et al. Phys. Rev. Lett, {\bf 83%
} 5090 (1999)

\bibitem{cit-marcus-def-dot}  M. Switkes, C. M. Marcus, K. Campman and A. C.
Gossard, Science 283, 1905 (1999).

\bibitem{cit-Vergini-Saraceno}  E. \ Vergini and M. Saraceno, Phys. Rev. E
{\bf 52}, 2204 (1995).

\bibitem{dcohenannal}
D. Cohen, Phys.Ann. Phys. (N.Y.) {\bf 283}, 175 (2000).

\bibitem{band1}
A. A. Gribakina, V. V. Flambaum, and G. F. Gribakin, Phys. Rev. E {\bf 52}, 5667 (1995).

\bibitem{band2}
W. Wang, F. M. Izrailev, and G. Casati, Phys. Rev. E {\bf 57} 323 (1998).


\bibitem{dcohenpara}
D. Cohen, A. Barnett and E.J. Heller, Phys. Rev. E {\bf 63}, 46207 (2001).

\bibitem{Cohen-Heller} D. Cohen and E. Heller, Phys. Rev. Lett. 
{\bf 84} 2841 (2000).

\bibitem{wisniacki4}
D.A. Wisniacki, unpublished.

\bibitem{karkuszewski} Z. P. Karkuszewski, C. Jarzynski and W. H. Zurek,
Phys. Rev. Lett. 89, 170405 (2002).

\bibitem{jordan} A. Jordan and M. Srednicki,
quant-ph/0112139.

\bibitem{note} Our system do not exhibit such a region due to
inclusion of the cut-off function. This region can be observed
for large enough $\delta x$ but in this case we can not deal 
with the numerics.  

\end{thebibliography}
\end{document}